\documentclass[aps,prb,twocolumn,showpacs,groupedaddress]{revtex4}

\usepackage{amsmath}
\usepackage{amssymb}
\usepackage{graphicx}
\usepackage{dcolumn}
\usepackage{bm}
\usepackage{subfigure}

\usepackage[normalem]{ulem}

\begin{document}

\title{Two-dimensional molecular para-hydrogen and ortho-deuterium at 
zero temperature}

\author{C. Cazorla$^{1,2,3}$ and J. Boronat$^{4}$}
\affiliation{
$^{1}$London Centre for Nanotechnology, UCL, London WC1H OAH, UK \\
$^{2}$Department of Physics and Astronomy, UCL, London WC1E 6BT, UK \\
$^{3}$Materials Simulation Laboratory, London WC1E 6BT, UK \\
$^{4}$Departament de F\'isica i Enginyeria Nuclear, UPC, Campus Nord B4-B5, 
E-08034 Barcelona, Spain
}
\

\begin{abstract}

We study molecular para-hydrogen (p-${\rm H_{2}}$) and ortho-deuterium
(o-${\rm D_{2}}$)  in two dimensions and in the limit of zero temperature
by means of the diffusion Monte Carlo method.  We report energetic and
structural properties of both systems like the total and  kinetic energy
per particle, radial pair distribution function, and  Lindemann's ratio  in
the low  pressure regime. By comparing the total energy per particle as a
function  of the density in liquid and solid p-${\rm H_{2}}$~, we show that
molecular para-hydrogen, and also ortho-deuterium,  remain solid at zero
temperature. Interestingly, we assess the quality of three different
symmetrized  trial wave functions, based on the Nosanow-Jastrow model, in
the p-${\rm H_{2}}$ solid  film at the variational level.  In particular,
we analyze a new type of symmetrized trial wave function which has been
used very recently to describe  solid $^{4}$He  and found that also
characterizes hydrogen satisfactorily.  With this wave function, we show
that  the one-body density  matrix $\varrho_{1} (r)$ of solid p-${\rm
H_{2}}$  possesses off-diagonal long range order, with a condensate fraction
that increases sizably in the negative pressure regime.

\end{abstract}

\pacs{61.50.Ah, 67.70.+n, 67.80.-s, 67.90.+z}

\maketitle

\section{Introduction}

Quantum crystals like helium and hydrogen are intriguing systems of 
fundamental physical interest. Due to the light mass of their constituents
and relatively weak interparticle attraction, quantum solids exhibit large 
kinetic energy and Lindemann's ratio even in the limit of zero temperature
($T \sim 10^{0}-10^{-3}$~K).  In consequence, anharmonic effects and atomic
quantum exchanges are of importance in this class of crystals. 
Moreover, in the last few years a series of ultra-low temperature experiments
performed in solid $^{4}$He by different groups has led to a renewed
interest on the possibility of superfluidity and/or Bose-Einstein
condensation (BEC)  in quantum
solids.~\cite{chan04a,chan04b,sasaki06,rittner06}  Essentially, these
experiments analyze the quantum behavior of the helium
crystal upon rotation  or seek for some thermodynamic and/or structural
anomaly signalizing a possible normal-to-superfluid phase transition.
Despite that most of the observations fairly agree in locating the onset of
superfluidity ($75 - 150$~mK), there is a large dispersion in the value of
the measured superfluid fraction   $\rho_{s}/\rho$ ($\simeq 0.3 -
0.005$~\%).  At present, there is lack of conclusive arguments for
explaining these discrepancies but it   is widely accepted that the purity of
the sample and the presence of crystalline defects play on it a relevant 
role.~\cite{rittner07,chan08}    On the other hand, there is overall
agreement among  microscopic full quantum calculations in practically
ruling out superfluidity in the perfect  (free of defects) bulk
configuration. It is worth noticing that  usual techniques devised to
study classical crystals (that is crystals composed  of heavier elements
and with larger cohesive energies), like for instance harmonic based
approaches, are not longer suitable for quantum solids and calculations on
them are in most cases challenging.~\cite{cazorla08a,cazorla08b}  

In the present work, we present a theoretical study of two-dimensional (2D)
molecular 
para-hydrogen (p-${\rm H_{2}}$) and ortho-deuterium (o-${\rm D_{2}}$) at
zero temperature  by means of the diffusion Monte Carlo (DMC)
method~\cite{hammond94,guardiola98,ceperley79} and the semi-empirical
radial pair  interaction due to Silvera and Goldman.~\cite{silvera78}
Hydrogen is a very interesting
and challenging system  which has been investigated very intensively during
the last half century.  As a matter of fact, hydrogen is the most abundant
element in  the universe and from a technological point of view it is
considered among the most promising green combustibles of the near future.
Very interestingly, hydrogen has been predicted  to exhibit a new state of
matter at very high pressures ($P \sim 400$~GPa) in which superfluidity
and  superconductivity might coexist.~\cite{ashcroft04,ashcroft05} 

In this work, we restrict our analysis of molecular hydrogen and deuterium to the low
pressure regime  ($P \sim 0$) and zero temperature. Contrarily to what
occurs in helium, molecular hydrogen freezes at a temperature of 
13.96 K  in spite of its lighter mass, given that the interactions between
particles are more  attractive (the minimum of the interaction between
hydrogen molecules amounts to $\sim -37$~K while in helium  is $\sim
-10$~K).  One of our motivations for carrying out the present study
was to unravel whether liquid p-${\rm H_{2}}$  could be stabilized or not
at zero temperature by reducing the dimensionality  with respect to
the bulk.  This possibility appears to be very appealing since it
would provide a chance for  superfluidity and Bose-Einstein condensation
(BEC) to be observed in a quantum liquid different from helium. In fact,
p-${\rm H_{2}}$ in one dimension and inside a carbon nanotube has already
been studied in the zero-temperature limit and predicted to be liquid at
its equilibrium density.~\cite{gordillo00} Also, small drops with a number
of molecules  $N \alt 26$ present superfluid
character.~\cite{drops1,drops2,drops3} On the other hand, it has been reported
recently an experiment
performed on molecular ortho-deuterium pre-plated on krypton at very 
low temperatures ($T \sim 1$~K) in which
it is claimed evidence for the existence of a  reentrant o-${\rm D_{2}}$
liquid phase.~\cite{wiechert04} As it will be presented in short, our
results show that no first order solid-liquid phase  transition
occurs in two-dimensional ${\rm H_{2}}$ at zero temperature. On account of
this result,  we straightforwardly reject this possibility also for o-${\rm
D_{2}}$ since deuterium  molecules are heavier and their intermolecular 
 interactions are considered equal to the H$_2$-H$_2$ ones. Consequently, most of
the effort done in this work has been devoted to achieve an accurate
description of the ground state of two-dimensional solid p-${\rm H_{2}}$
and o-${\rm D_{2}}$~.

In this work, we calculate by means of the diffusion Monte Carlo method
(DMC) some energetic and structural properties of both hydrogen and
deuterium films near equilibrium. Quantities like the kinetic energy per
particle and Lindemann's ratio have been computed within the pure
estimator  approach~\cite{liu74,reynolds86,casulleras95} in order to remove
any possible bias coming out from the trial wave function used for
importance sampling. In this way, we quote quantum isotopic effects in
hydrogen directly and only within the statistical uncertainty. Previous to
the DMC results, we present  a variational Monte Carlo (VMC) study of  the
p-${\rm H_{2}}$ crystal in which  we have tested the quality of several
symmetrized and unsymmetrized trial wave functions.  With this analysis, we
determine the effect of symmetrization on the total energy and the
relevance of molecule exchanges along the simulation. Moreover, we analyze
which symmetrized wave functions can be implemented in DMC to the end of
estimating the possible  superfluidity and Bose-Einstein condensation (BEC)
of the solid with simultaneous accurate description of their energetic
and structural properties.  In particular, we have studied in detail a
symmetrized trial wave function, named $\psi_{JG}^{S3}$  in this work (the
notation will become clear later),  which has been proposed and used
recently  to study bulk solid $^{4}$He.~\cite{cazorla08c}  Here, we find
that $\psi_{JG}^{S3}$ also characterizes solid hydrogen in two dimensions
accurately.    

Interestingly, we assess the behavior of the
one-body density matrix $\varrho_{1} (r)$ of p-${\rm H_{2}}$ with  density
by means of the  symmetric trial wave function $\psi_{JG}^{S3}$. In all
cases, a very small condensate fraction $n_0$ is observed. For densities below
the equilibrium one, and near the spinodal point, a significant
increase of $n_0$  is observed pointing to the emergence of a finite
superfluid density.

This paper is organized as follows. In Sec.~\ref{sec:techniques}, we
present a brief description of  the semiempirical pair potential and
techniques used throughout this work. Next, in Sec.~\ref{sec:h2} and
\ref{sec:d2},  we report our variational and diffusion Monte Carlo results
for p-${\rm H_{2}}$ and  o-${\rm D_{2}}$ in two dimensions, respectively.
Sec.~\ref{sec:possiblesuper} is devoted to the examination of the one-body
density matrix $\varrho_{1} (r)$ obtained with $\psi_{JG}^{S3}$ and its
dependence with the density. Finally, in the Sec.~\ref{sec:discussion} we summarize 
the main results presented in this work.

\section{Molecular interaction and method}
\label{sec:techniques}

The ${\rm H_{2}}$ (${\rm D_{2}}$) molecule is composed of two hydrogen
(deuterium) atoms  linked by a covalent bond, which in the para-hydrogen
(ortho-deuterium) state   possesses spherical symmetry (total angular
momentum zero).  The energy scale involved in electronic excitations   
($\sim 10^{5}$ K) is orders of magnitude larger than the intermolecular
one  ($\sim 10^{1}$ K), thus to model the ${\rm H_{2}-H_{2}}$ (or ${\rm
D_{2}-D_{2}}$) interaction by means of a radial  pair-potential and
consider the molecules as point-like turns out to be justified upon the
condition  of low or moderate pressures.  In this work, we have adopted the
well-known and commonly used semiempirical Silvera-Goldman pair 
potential.\cite{silvera78} 
This potential has proved to perform excellently at low  temperature and
at the pressure regimes in which we  are interested.    

The ground state of para-H$_2$ and ortho-D$_2$ is determined using the DMC
method.
DMC is a zero-temperature method which provides the exact ground-state
energy of many-boson interacting systems within some statistical
uncertainty.~\cite{hammond94,guardiola98,ceperley79}  This technique
is based on a short-time approximation for the Green's function
corresponding  to the imaginary time($\tau$)-dependent Schr\"{o}dinger equation.
Despite this method is algorithmically simpler
than domain Green's function Monte Carlo,~\cite{ceperley79,kalos74} it
presents some $\left(\Delta \tau\right)^{n}$ bias coming from the
factorization of the imaginary time propagator
$e^{-\frac{\Delta\tau}{\hbar}{\rm H}}$. Our implementation of DMC is
quadratic,~\cite{chin90} hence the control of the time-step bias is
efficiently controlled   given that the required $\Delta\tau \to 0$
extrapolation is nearly eliminated by choosing a sufficiently small time
step. The Hamiltonian ${\rm H}$, describing our system is 
\begin{equation}
\label{eq:hamiltonian}
{\rm H} = - \frac{\hbar^{2}}{2m} \sum_{i=1}^{N} \nabla^{2}_{i} +
       \sum_{i<j}^{N} V(r_{ij})~,
\end{equation}  
and  the corresponding
Schr\"odinger equation in imaginary time ($it \equiv \tau$), 
\begin{equation}
\label{eq:schrodinger}
-\hbar\frac{\partial \Psi({\bf R},\tau)}{\partial \tau}= 
\left({\rm H}-E\right)\Psi({\bf R},\tau) \ , 
\end{equation}
with $E$ an arbitrary constant.  Equation (\ref{eq:schrodinger}) 
can be formally solved by expanding the
solution $\Psi({\bf R}, \tau)$ in the basis set of the energy
eigenfunctions $\{\Phi_{n}\}$ (${\bf R}\equiv \{ {\bf r}_{1}, {\bf
r}_{2},\ldots,{\bf r}_{N} \}$). It turns out that $\Psi({\bf R}, \tau)$
tends to the ground-state wave function $\Phi_{0}$ of the system for an
infinite imaginary time as well as the expected value of the Hamiltonian
tends to the ground-state value $E_{0}$.  The hermiticity of the
Hamiltonian guarantees the equality   
\begin{equation}
\label{eq:groundstate1}
E_{0} = \frac{\left<\Phi_{0}|{\rm H}|\Phi_{0}\right>}{\left<\Phi_{0}|\Phi_{0}\right>}=
\frac{\left<\Phi_{0}|{\rm H}|\psi_{T}\right>}{\left<\Phi_{0}|\psi_{T}\right>} = \langle {\rm H} \rangle_{DMC} ~, 
\end{equation}
where $\psi_{T}$ is a convenient trial wave function. 
As a consequence, the ground-state energy of the system can be computed by 
calculating the integral
\begin{equation}
\label{eq:integral}
\langle {\rm H} \rangle_{DMC}= \lim_{\tau \to\infty} 
\int_{V} E_{L}\left({\bf R}\right) f\left({\bf R},\tau\right) d{\bf R} \quad ,
\end{equation} 
where $f\left({\bf R},\tau\right)=\Psi\left({\bf
R},\tau\right)\psi_{T}\left({\bf R}\right)$, and $E_{L}\left({\bf
R}\right)$ is the local energy defined as $E_{L}({\bf R})= {\rm
H}\psi_{T}\left({\bf R}\right)/\psi_{T}\left({\bf R}\right)$.  The
introduction of $\psi_{T}\left({\bf R}\right)$ in $f\left({\bf
R},\tau\right)$ is known as importance sampling and its use is important to
reduce the variance of Eq. (\ref{eq:integral}) to a manageable level   
(for instance, by imposing $\psi_{T}\left({\bf R}\right)=0$ when
$r_{ij}$ is smaller than the core of the pair interaction).

In this work, all the operators diagonal in real-space which do not commute
with the Hamiltonian,  that is $[ {\rm H}, \hat{O} ] \neq 0$, have been
sampled using the pure estimator
technique based on forward walking.~\cite{liu74,reynolds86,casulleras95} 
Essentially, with this
method the possible bias introduced by $\psi_{T}$ in the mixed  estimator
$\left< \Phi_{0} | \hat{O} | \psi_{T}\right>$ are removed by proper
weighting of the  configurations generated along the simulation.

\section{Molecular para-hydrogen}
\label{sec:h2}

\subsection{Variational Monte Carlo results}
\label{subsec:vmc}

In this section, we present a variational Monte Carlo (VMC)
study of two-dimensional p-${\rm H_{2}}$   which provides us with the most
convenient trial wave function (twf) to be used in subsequent DMC
calculations and valuable physical insight on the system itself. In brief, 
the VMC method relies on the variational principle which  states that given
a Hamiltonian the energy difference $E_{L} - E_{0}$ averaged over the
probability density distribution  $|\psi_{T}|^{2}$ is always positive and
it decreases as the overlapping between $\psi_{T}$ and the true
ground-state wave function increases ($E_L$ and $E_0$ are the local energy
defined in the previous section and the ground-state energy, respectively).  
In the present work, the main
variational effort has been devoted to achieve an accurate description of
the 2D solid phase. We have checked by means of VMC and DMC that in both
p-${\rm H_{2}}$ and o-${\rm D_{2}}$ systems the triangular configuration is
the stable one at all the studied densities.   The energies reported in
this section have been calculated at the density $\rho =
0.060$~\AA$^{-2}$~,  which corresponds to the variational equilibrium density 
of the solid.

In order to determine the nature of the ground-state of the system we have
also carried out simulations for the liquid phase. In this case, 
the trial wave function is of Jastrow type, 
\begin{equation} 
\label{eq:twfliquid} 
\psi_{J}({\bf
r}_{1},{\bf r}_{2},\ldots,{\bf r}_{N})=\prod_{i<j}^{N}{\rm f}_{2}(r_{ij})~,
\end{equation} 
where the two-body factors ${\rm f_{2}}$ account for the
molecular correlations arising in the system  due to 
pair interactions. These two-body correlation factors have been chosen of 
McMillan  form, ${\rm f}_2 = e^{-\frac{1}{2}\left( \frac{b}{r}\right)^{5}}$~, and
as best value of the variational parameter $b$ we obtain $3.70$~\AA~.

\begin{figure}
\centerline{
        \includegraphics[width=0.8\linewidth]{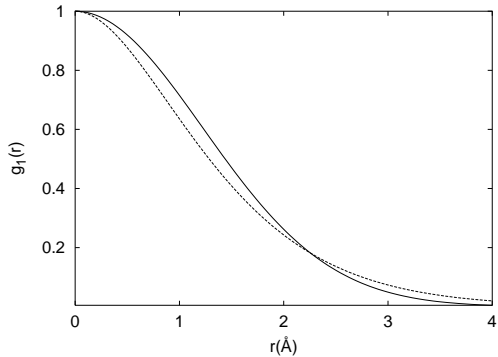}}%
        \caption{Optimized Gaussian and Pad\'e functions (solid and dashed line, respectively)
for solid p-${\rm H_{2}}$ at the density $\rho = 0.060$\AA$^{-2}$.
The decay of ${\rm g}_{P}(r)$ to zero is smoother than that of 
${\rm g}_{G}(r)$, and in the region around the origin ${\rm g}_{P}(r)$ is narrower.}
\label{fig:onebody}
\end{figure}

For the solid phase, an additional one-body factor is introduced \emph{ad
hoc} in the trial wave  function to the end of reproducing the periodic
order of the system and so making the sampling over the space of
configurations more efficient (Nosanow-Jastrow model).  Such one-body
factor consists in a productory of localizing functions centered on the
positions that   define the perfect crystal configuration (sites), given by
the family of vectors $\lbrace {\bf R}_{i}\rbrace$~ 
\begin{equation}
\label{eq:trialwvfuncsol}
\psi_{NJ}({\bf r}_{1},{\bf r}_{2},\ldots,{\bf r}_{N})=
\psi_{J}\prod_{i}^{N}{\rm g}_{1}(|{\bf r}_{i}-{\bf R}_{i}|)~. 
\end{equation}
We have explored two different trial wave functions based on
$\psi_{NJ}$~, each one consisting in a different choice   for ${\rm
g}_{1}(r)$. The first model corresponds to the  standardly used Gaussian
function, 
\begin{equation}
\label{eq:gauss}
{\rm g}_{G}(r) =\exp{\left(-\frac{a_{G}}{2}r^{2}\right)}~,
\end{equation}
while the second is a Pad\'e function, defined as 
\begin{equation}
\label{eq:pade}
{\rm g}_{P}(r)=\exp{\left(-\frac{a_{P} r^{2}}{1 + c_{P} r}\right)}~,
\end{equation}
where $a_{G}$, $a_{P}$ and $c_{P}$ are variational parameters to be
optimized. Even though  ${\rm g}_{G}(r)$ and ${\rm g}_{P}(r)$ are
analytically quite similar, the asymptote of ${\rm g_{G}}$  can be chose so
as to decay to zero less abruptly than that of ${\rm g}_{G}(r)$ (see
Fig.~\ref{fig:onebody}).  This feature can be used in the simulations 
for increasing somewhat the degree of delocalization of the
molecules at the expense, however, of an increase of the kinetic energy
within the surroundings of the equilibrium positions (where ${\rm g}_{P}(r)$
varies more rapidly than ${\rm g}_{G}(r)$~).  
In Table I~, we report the best energies 
obtained in the optimization process of    $\psi_{NJ}$ with
Gaussian and Pad\'e functions for ${\rm
g}_{1}(r)$~ as well as the optimal set of parameters.  
As one can see, in both cases the lowest energy
obtained is $-21.3(1)$~K~. It is worthwhile noticing that when the asymptotes
of the Pad\'e factors are widened (that is, the value of $c_{P}$ is
increased), or equivalently, when the molecules are left to move more
freely around the equilibrium  positions, the energy
of the system increases. Given the variational equivalency between ${\rm
g}_{G}(r)$ and ${\rm g}_{P}(r)$~, we opt for $\psi_{NJ}$ with Gaussian factors 
and optimal parameters $b_{G} = 3.45$~\AA and $a_{G} = 0.67$\AA$^{-2}$ in
our subsequent DMC calculations.

It is well known that $\psi_{NJ}$ is not symmetric under the permutation of
particles  (that is, $\psi_{NJ}({\bf r}_1,{\bf r}_2,\ldots,{\bf r}_N) \neq
\psi_{NJ}({\bf r}_2,{\bf r}_1,\ldots,{\bf r}_N)$). This property is manifestly
incorrect for a system of indistinguishable bosons.  Nevertheless, the use of the
Nosanow-Jastrow model is widely spread within the field of microscopic 
calculations since it is assumed that the effect of symmetrization on the
total and partial energies of quantum solids  is practically negligible. In
fact, we will show in brief that this is also the case for
two-dimensional hydrogen.   To this end, we have tested  the quality
of several symmetrized trial wave functions in the variational study of 2D solid 
p-${\rm H_{2}}$.  As mentioned in the Introduction, our
motivation for this analysis is twofold: 
estimate the influence in the energy of molecular quantum exchanges, and 
chiefly, study the possibility of its use as importance sampling in DMC to determine
the possible existence of Bose-Einstein condensation (BEC) and superfluidity.

\begin{table*}
\begin{center}
\label{tab:variational}
\begin{tabular}{c | c c c c | c c c c c}
\hline
\hline
    &  $ b_{G}$ (\AA)  &  $ a_{G}$(\AA$^{-2}$) &  $ E/N (K)$  & $ Q_{2}~(\%)$ &  $ b_{P}$ (\AA) &
$ a_{P}$ (\AA$^{-2}$)  &  $ c_{P}$(\AA$^{-1}$) & $ E/N (K)$ & $ Q_{2}~(\%)$  \\
\hline
$\psi_{NJ}$ & $ 3.45  $  &   $ 0.67  $    &   $ -21.3(1) $ &     & $3.32$  & $  0.46  $  &  $  0.20  $    &    $ -21.3(1)$ &    \\
$\psi_{JL}^{S1}$ & $ 3.45  $  &   $ 0.67  $  & $-21.3(1)$ & $ 4\cdot 10^{-3} $ &  $3.32$  & $  0.46  $  &  $  0.20  $ & $-21.3(1)$ &  $ 6\cdot 10^{-2} $ \\
$\psi_{JL}^{S2}$ & $  3.58 $   &   $0.38$  & $-17.9(1)$ &  &  $3.70$ &  $0.69$ &  $0.39 $ & $-17.9(1)$ &   \\
$\psi_{JG}^{S3}$ & $ 3.45  $   &   $ 0.61  $  &  $-20.4(1)  $ &   &   &   &   &    &    \\  
\hline
\hline
\end{tabular}
\end{center}
\caption{Optimal variational total energy per particle obtained at the 
density $\rho$= 0.060\AA$^{-2}$~
with different trial wave function models.
Values appearing on the left(right)-side of the table correspond to
one-body factors ${\rm g}_{1}(r)$ adopted in the form
${\rm g}_{G}(r)$~(${\rm g}_{P}(r)$)~.}
\end{table*}

We have studied three different symmetrized trial wave functions.
The first model consists in a permanent of monoparticular functions 
containing the $N!$ possible permutations $\lbrace P \rbrace$ of the $N$ 
particles among the different lattice sites,  expressed as
\begin{equation}
\label{eq:sym2}
\psi^{S1}_{JL}({\bf r}_{1},{\bf r}_{2},\ldots,{\bf r}_{N})=
\psi_{J}\sum_{\lbrace P \rbrace} \prod_{i=1}^N {\rm g}_{1}\left({\bf r}_{i}-
{\bf R}_{Pi}\right) .
\end{equation}
Due to the algebraic difficulties arising in the implementation of 
permanents (contrarily to what occurs with determinants), the sampling of 
$\psi^{S1}_{JL}$ must be divided into two different parts, one performed in
the space of spatial configurations and the other in the space of
permutations.~\cite{ceperley78} The acceptance probability for a proposed
change of position of the particle labeled $i$, ${\bf r}_{i}\to{\bf
r}'_{i}$, corresponds to 
\begin{equation}
\label{eq:acc1}
q = \min \left(1,\frac{\psi_J({\bf r}')^{2}{\rm g}_{1}({\bf r}'_{i}-{\bf
R}_{i}){\rm g}_{1}({\bf r}'_{i}-{\bf R}_{Pi})}
{\psi_J({\bf r})^{2}{\rm g}_{1}({\bf r}_{i} - {\bf R}_{i}){\rm g}_{1}({\bf
r}_{i}-{\bf R}_{Pi})}\right) ,
\end{equation}
where the subindex $Pi$ can take any of the $N$ possible lattice
sites. On the other side, the acceptance probability for a
proposed site permutation between the $i$ and the $j$ particles, ${\bf
R}_{Pi} \leftrightarrow {\bf R}_{Pj}$, is 
\begin{equation}
\label{eq:acc2}
Q_{2} = \min \left(1,\frac{{\rm g}_{1}({\bf r}_{j} - {\bf R}_{Pi}) {\rm
g}_{1}({\bf r}_{i} - {\bf R}_{Pj})}
{{\rm g}_{1}({\bf r}_{j} - {\bf R}_{Pj}){\rm g}_{1}({\bf r}_{i} - {\bf R}_{Pi})}\right)~.
\end{equation}
Notice that permutations involving more than two particles are not sampled
since the acceptance level for swap permutations is already extremely low.

The optimal results obtained with $\psi^{S1}_{JL}$, using Gaussian and Pad\'e
${\rm g_{1}}(r)$ functions, are reported in
Table~I~. By comparing the variational energies obtained with 
$\psi_{NJ}$ and $\psi^{S1}_{JL}$, we show that symmetrizing $\psi_{NJ}$ 
with the above prescription has not appreciable effects on the total energy of
p-${\rm H_{2}}$. Nevertheless, it must be said that one should not draw
other conclusive statements about the effects of a full symmetrization just
based on the approximation of the permanent by a reduced sampling in the permutation 
space of the type (\ref{eq:acc2})~. In fact, the acceptance rate of permutations is 
so low (column $Q_{2}$ in Table~I~) that sampling $\psi^{S1}_{JL}$ 
efficiently turns out to be quite challenging.

\begin{figure}
\centerline{
        \includegraphics[width=0.8\linewidth]{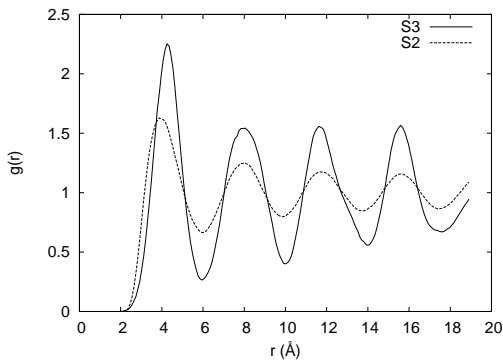}}%
        \caption{Variational radial pair distribution function $g(r)$ of two-dimensional
molecular hydrogen at the density $\rho = 0.060$\AA$^{-2}$ obtained with twf $\psi^{S2}_{JG}$
and $\psi^{S3}_{JG}$~.}
\label{fig:grsym}
\end{figure}

The second model of symmetrized trial wave function $\psi^{S2}_{JL}$ 
consists in a productory of sums in the form
\begin{equation}
\label{eq:sym1}
\psi^{S2}_{JL}\left({\bf r}_{1}, {\bf r}_{2},\ldots,{\bf r}_{N}\right)=\psi_{J}\prod_{i=1}^N \left(\sum_{j=1}^N
{\rm g}_{1}\left({\bf r}_{i}- {\bf R}_{j}\right) \right)~. 
\end{equation}
This trial wave function has been proposed very recently by Zhai and Wu~\cite{zhai05} 
and has been suggested to be of possible relevance for the study of the supersolid.
In fact, $\psi^{S2}_{JL}$ avoids any explicit sampling in permutation space hence 
turns out to be well-suited for being used as importance sampling in DMC simulations. 
However, as one can see in Table~I the best variational energy obtained
with this model is sizably larger than the ones obtained with
$\psi_{NJ}$  and $\psi_{JL}^{S1}$, in  both Gaussian and Pad\'e cases. In
fact, the variational energy obtained with $\psi^{S2}_{JL}$ is very similar to the
one calculated with $\psi_{J}$ for the liquid ($\sim -17.4$~K).
In doing the simulation with this wave function it is observed that
particles diffuse excessively within the container giving place to glassy-like configurations; 
we have checked this feature by monitoring the radial pair
distribution function (see Fig.~\ref{fig:grsym}) and mean squared
displacement (which grows steadily with time).
The reason for this excessive atomic diffusion is that the way in which the one-body
factor is symmetrized in $\psi^{S2}_{JL}$ does not
penalize  multiple occupation of a same lattice site. This feature will be
illustrated in short by means of a simple example involving two particles
moving in one dimension. Moreover, if the width of ${\rm g_{1}}(r)$ 
is narrowed in order to avoid such unrealistic molecular diffusion, the total energy of
the system is worsened because of the rapid increase of kinetic energy.

The third type of symmetrized trial wave function reads 
\begin{equation}
\label{eq:sym3}
\psi^{S3}_{JG}\left({\bf r}_{1}, {\bf r}_{2},\ldots,{\bf r}_{N}\right)=\psi_{J}\prod_{j=1}^N \left(\sum_{i=1}^N
{\rm g}_{1}\left({\bf r}_{i}- {\bf R}_{j}\right) \right) ,
\end{equation}
and it is also straightforward to implement in DMC codes. This type of
trial wave function has been proposed very recently by Cazorla \emph{et
al.}~\cite{cazorla08c} and has been used to set an upper bound of $10^{-5}$
for the superfluid fraction of perfect  crystalline bulk $^{4}$He at zero
temperature.  $\psi_{JG}^{S3}$ and $\psi_{JG}^{S2}$ look
similar, the difference being that the productory and summatory in
$\psi_{JG}^{S3}$ run over sites and particles, respectively, while in
$\psi_{JG}^{S2}$ is the other way around. However,  $\psi^{S3}_{JG}$ and
$\psi^{S2}_{JG}$ lead to completely different variational energies  (see
Table~I~). The best variational result obtained with 
$\psi^{S3}_{JG}$
amounts to $-20.4(1)$~K, which is only $0.9$~K larger than the one 
calculated with $\psi_{NJ}$ or $\psi^{S1}_{JL}$ but $2.5$~K smaller than
the one corresponding to $\psi^{S2}_{JG}$. In this case, the radial pair
distribution function and mean squared displacement    follow typical
solid-like patterns (see Fig.\ref{fig:grsym}).

\begin{figure}
\centerline{
        \includegraphics[width=0.8\linewidth]{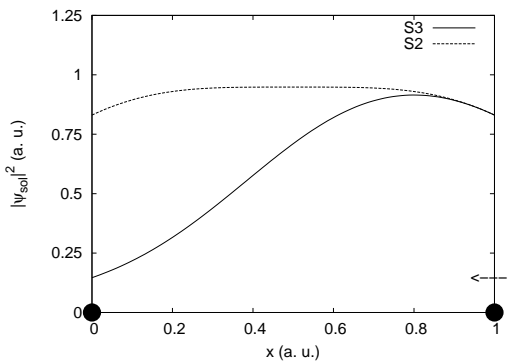}}%
        \caption{Squared $\psi^{S2}_{JG}$ and $\psi^{S3}_{JG}$ (with
	$\psi_J =1$) 
                 in the simple case of two particles moving in one dimension and
                 sites separated by one arbitrary unity.}
\label{fig:symmetrized}
\end{figure}

Contrarily to what occurs
with $\psi^{S2}_{JG}$, the multiple occupation of a same site is now
penalized by the wave function and hence crystal order is sustained. 
To the end of illustrating this feature, which appears to be the main
difference between $\psi^{S2}_{JG}$ and $\psi^{S3}_{JG}$, we have
analyzed the simple case of two particles moving  in a one-dimensional
lattice. For the sake of simplicity, we have assumed that the distance 
between the equilibrium positions of the particles is one, that the
parameter entering the Gaussian factors in Eq.(\ref{eq:sym2}) and
(\ref{eq:sym3}) is $a = 1/2$, in arbitrary units (a.u.), and that the Jastrow
factor is switched off ($\psi_J=1$). The value of the
squared wave function for $\psi^{S2}_{JG}$ and
$\psi^{S3}_{JG}$, $|\psi_{sol}|^{2}$, obtained by keeping fixed one of the
particles in the site located at the origin and then moving the other
particle towards it,  is plotted in Fig.~\ref{fig:symmetrized} for the
interval $0 \leq x \leq 1$~.  As one observes in there, the value of
$\psi^{S2}_{JG}$ at $x = 1$ and $0$ (that is, each particle is  placed over
one site or both are at the same position, respectively) is the same, 
whereas $\psi^{S3}_{JG}(x = 1) > \psi^{S3}_{JG}(x = 0)$.  
This effect is what we have previously refereed to
as ``penalized by the trial wave function". It is also noted that the
value of $\psi^{S2}_{JG}$ is maximum at half the way between $0$ and $1$,
not so for $\psi^{S3}_{JG}$, hence $\psi^{S2}_{JG}$ will always promote
larger diffusion of the molecules.

\subsection{Diffusion Monte Carlo results}
\label{subsec:DMC}

We have studied the energetic and structural properties of p-${\rm H_{2}}$
using the DMC method and $\psi_{NJ}$ (\ref{eq:trialwvfuncsol}) as trial
wave function. We have verified that the DMC energy and diagonal properties
obtained with $\psi_{NJ}$ are statistically indistinguishable from the ones
obtained using the symmetric wave function $\psi^{S3}_{JG}$
(\ref{eq:sym3}). The results presented in this section have been obtained for a 
rectangular plane containing 90 particles  with periodic boundary conditions 
in the two spatial directions. Internal parameters of the
simulations, namely the averaged population of walkers and time  step, are
$250$ and $5 \cdot 10^{-4}$~K$^{-1}$, respectively; these parameters have
been adjusted in order to reduce any possible bias to the level of the statistical
uncertainty ($\sim 0.05$~K).

\begin{figure}
\centerline{
        \includegraphics[width=0.8\linewidth]{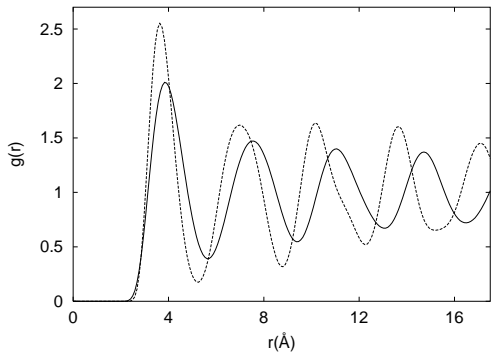}}%
        \caption{Radial pair distribution functions of 2D solid o-${\rm D_{2}}$ at the equilibrium
         density $\rho_{0} = 0.078$~\AA$^{-2}$ (dashed line), and of p-${\rm H_{2}}$ at the
         density $0.068$~\AA$^{-2}$ (solid line).
		 }
\label{fig:deuteriumgr}
\end{figure}

In Table~II~, we report the total ground-state energy per particle, $E/N$,
corresponding to   2D solid p-${\rm H_{2}}$ at some densities. The
pure (unbiased) estimation of the potential, $V/N$, and
kinetic energies, $T/N$, are also quoted therein. The energy
results have been corrected for the finite size of the
simulation plane  by assuming the radial pair distribution function $g(r)$
to be one beyond the distance $R_{\rm max}=L/2$, with $L$ being the size of the
plane.
Assuming $g(r) \sim 1$ beyond $R_{\rm max}$ could seem a crude approximation for
crystals since this function shows periodic structure (see for instance
Fig.~\ref{fig:deuteriumgr}). 
However, the
periodic oscillations of $g(r)$ around unity might suggest that in average
this  approximation is essentially correct.  In order to test the
reliability of this finite size correction we have  
calculated the total energy per particle in a plane containing $90$~, $120$
and $168$ molecules  at the density $\rho = 0.0597$~\AA$^{-2}$~; we obtain
$E/N = -22.19(2)$, $-22.16(2)$ and $-22.15(2)$~K, respectively,  thus
achievement of convergence within the present statistical uncertainty is proved.  

\begin{table}
\label{tab:results}
\begin{center}
\begin{tabular}{ c c c c }
\hline
\hline
$ \rho $(\AA$^{-2}$)   &  $E/N$ (K)   &  $V/N$ (K)   &  $T/N$ (K)  \\
\hline
$ 0.053 $   &  $  -19.42(2) $  &  $ -35.73(3)  $  &  $ 16.30(3)$\\
$ 0.060 $   &  $  -22.21(2) $  &  $ -43.67(3)  $  &  $ 21.46(3)$\\
$ 0.065 $   &  $  -23.27(2) $  &  $ -49.23(4)  $  &  $ 25.96(4)$\\
$ 0.067 $   &  $  -23.42(2) $  &  $ -51.68(4)  $  &  $ 28.26(4)$\\
$ 0.070 $   &  $  -23.19(2) $  &  $ -54.83(4)  $  &  $ 31.64(4)$\\
$ 0.076 $   &  $  -21.30(2) $  &  $ -59.22(5)  $  &  $ 37.92(5)$\\
$ 0.083 $   &  $  -14.23(2) $  &  $ -62.40(7)  $  &  $ 48.17(7)$\\
\hline
\hline
\end{tabular}
\end{center}
\caption{Ground-state energy $E/N$, potential energy $V/N$, and kinetic energy $
T/N$, per particle of solid 2D p-${\rm H_{2}}$. Potential and kinetic energies are 
obtained with pure estimators.
}
\end{table}

\begin{figure}
\centerline{
        \includegraphics[width=0.8\linewidth]{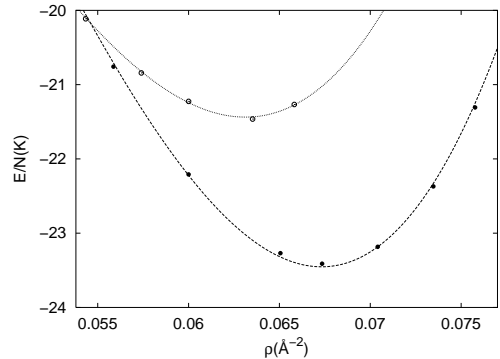}}%
        \caption{ Total ground-state energy per particle of liquid (dotted line) 
	and solid (dashed line) 2D p-${\rm H_{2}}$ at zero
temperature. The lines correspond to polynomial fits of our results (empty and 
filled circles);
the statistical errors bars are smaller than the symbol size.}
\label{fig:eos}
\end{figure}

The energy per particle corresponding to liquid and solid 2D p-${\rm
H_{2}}$ at zero temperature is plotted in Fig.~\ref{fig:eos} as a function
of the density. The simulation of the metastable liquid phase uses a Jastrow 
wave function $\psi_{J}$ (\ref{eq:twfliquid}) as importance sampling.
The lines in Fig.~\ref{fig:eos} correspond to  polynomial fits to our results in the form
\begin{equation}
\label{eq:efit}
E/N= e\left(\rho \right)= e_{0} + B\left(\frac{\rho-\rho_0}{\rho_0}\right)^2+
C\left(\frac{\rho-\rho_0}{\rho_0}\right)^3~.
\end{equation}
The pressure, compressibility and speed of sound (averaged for all the directions) 
are then easily derived from Eq.(\ref{eq:efit}) through the expressions,
\begin{equation}
\label{eq:pressfit}
P(\rho)=\rho^{2}\frac{\partial e(\rho)}{\partial \rho}
\end{equation}
\begin{equation}
\label{eq:comfit}
\kappa(\rho)=\frac{1}{\rho}\left(\frac{\partial\rho}{\partial P}\right)_{T}
\end{equation}
\begin{equation}
\label{eq:soundfit}
c(\rho)=\left(\frac{1}{m\kappa\rho}\right)^\frac{1}{2} .
\end{equation}
The optimal value of the parameters for the solid phase are $e_{0} =
-23.453(3)$~K,  $\rho_{0} = 0.0673(2)$~\AA$^{-2}$, $B = 121(2)$~K, and  $C =
152(8)$~K, where $e_{0}$ and $\rho_{0}$ are the  equilibrium energy per
particle and density, respectively. According to these figures, the
compressibility and speed of sound at the equilibrium density  are
$\kappa(\rho_0) = 0.0615(8)$~\AA$^{2}$/K and $c(\rho_0)= 998.6(1)$~m/s, 
the numbers quoted within parentheses being the statistical errors.
The equation of state of the liquid phase is also well described by the
polynomial form (\ref{eq:efit}) with optimal parameters 
$e_{0} =-21.43(2)$~K,  $\rho_{0} =0.0633(3)$~\AA$^{-2}$, $B = 75(7)$~K, and  $C =
69(9)$~K~.

Another magnitude of interest in the study of bulk systems is the spinodal
density $\rho_{S}$~. $\rho_{S}$ sets the limit for the system to remain in
a homogeneous phase since at this density   the compressibility grows to
infinite (or equivalently, the speed of sound becomes zero); in case of going below 
this point ($\rho < \rho_{S}$) the system breaks down into clusters.
According to our DMC calculations,  this low-limit density amounts to $\rho_{S}=
0.0548(1)$\AA$^{-2}$ in solid p-${\rm H_{2}}$.

A glance at Fig.~\ref{fig:eos} shows that the solid phase is the stable
one overall the  regime of positive pressures. Nevertheless, by looking at our 
results one could suggest a first order liquid-solid phase transition occurring  at
negative pressures, where the two equations of state cross each other.
Needless to be said, that this possibility deserves detailed exploration
since it could provide a chance for superfluidity to be observed in a
quantum liquid different from helium.   Aimed at this, we have simulated 2D
liquid p-${\rm H_{2}}$ down to densities of $0.039$\AA$^{-2}$ (the spinodal
density of the liquid is $\rho_{S} = 0.0519(1)$\AA$^{-2}$) and searched for
that transition by means of the Maxwell double-tangent construction.  Our
results show that a liquid-solid transition is not possible within the
range  set by the spinodal densities, and thus the possibility of liquid
p-${\rm H_{2}}$  in two dimensions must be rejected.

Our results for the equation of state of  p-${\rm H_{2}}$ can be compared
with two previous PIMC studies carried out on the same system. In
Ref~\onlinecite{gordillo97}, Gordillo and Ceperley obtained  $\rho =
0.064$~\AA$^{-2}$ for the equilibrium density of 2D solid p-${\rm H_{2}}$
at $ T = 1$~K; the authors of that work reported a figure with the energy
per particle as a function of the density, and the minimum of the curve is
located at $\sim -22.0$ K. A more systematic analysis of the same system
was performed later on by Boninsegni.~\cite{boninsegni04} In that  work,
the total energy per particle and  chemical potential are calculated at
several densities and within the temperature range $T=1-8$~K.  Subsequently,
an extrapolation of the low temperature results to absolute zero was
performed, leading to  $\rho_{0}^{\rm PIMC}=0.0668(5)$~\AA$^{-2}$,
$e_{0}^{\rm PIMC}=-23.25(5)$~K and $\rho_{S}^{\rm PIMC}=0.0585(10)$~\AA$^{-2}$.  We
note that the agreement between those zero-temperature extrapolated PIMC
values and our DMC results is fairly good,  specially in the case of the
equilibrium density $\rho_{0}$.

\begin{table}
\label{tab:lindemann}
\begin{center}
\begin{tabular}{c c c c c }
\hline
\hline
$ \rho$(\AA$^{-2}$)   &  $\gamma_{\rm H_{2}} $  & $ \zeta_{(10)}$   &  $\zeta_{(01)}$ &  $ \langle{\bf u^{2}}\rangle$(\AA$^{2}$) \\
\hline
$0.058$   &    $0.212(1)$  &  $0.00(2)$  &    $0.00(2)$  &      $0.19(1)$   \\
$0.060$   &    $0.197(1)$  &  $0.02(2)$  &    $0.00(2)$  &      $0.15(1)$   \\
$0.065$   &    $0.183(1)$  &  $0.02(2)$  &    $0.02(3)$  &      $0.12(1)$   \\
$0.067$   &    $0.178(1)$  &  $-0.01(1)$ &    $-0.02(2)$ &      $0.11(1)$   \\
$0.070$   &    $0.170(1)$  &  $-0.02(1)$ &    $0.00(2)$  &      $0.09(1)$   \\
$0.076$   &    $0.158(1)$  &  $0.00(2)$  &    $0.01(1)$  &      $0.07(1)$   \\
$0.083$   &    $0.146(1)$  &  $0.01(2)$  &    $0.01(1)$ &      $0.06(1)$   \\
\hline
\hline
\end{tabular}
\caption{Lindemann's ratio $\gamma_{\rm H_{2}}$, kurtosis $\zeta$, and mean squared displacement $\langle {\bf u^{2}}\rangle$
of 2D solid p-${\rm H_{2}}$ at different densities close to equilibrium (pure estimations).}
\end{center}
\end{table}

We have analyzed the structure of the 2D solid by calculating the radial 
pair distribution function $g(r)$,
\begin{equation}
\label{eq:grfunc}
g(r)=\frac{N-1}{\rho}\frac{\int|\Psi\left({\bf r}_{1},{\bf r}_{1}+{\bf
r},\ldots,{\bf r}_{N}\right)|^{2}
d{\bf r}_{1}d{\bf r}_{3} \ldots d{\bf r}_{N}}{ \int|\Psi\left({\bf
r}_{1},{\bf r}_{2},\ldots,{\bf r}_{N}\right)|^{2}d{\bf r}_{1} \ldots
d{\bf r}_{N}}~,
\end{equation}
and the Lindemann's ratio $\gamma_{\rm H_{2}}$,
\begin{equation}
\label{eq:lindemann}
\gamma =\frac{1}{a} \sqrt{\langle\frac{1}{N}\sum_{i=1}^N\left({\bf
r}_{i}-{\bf R}_{i} \right)^{2}\rangle}=
\frac{\langle {\bf u^2}\rangle^{\frac{1}{2}}}{a} ,
\end{equation}
where $a$ is the distance between nearest neighbors in the perfect
crystalline configuration. In Fig.~\ref{fig:deuteriumgr},
we plot $g(r)$ at the density $0.068$\AA$^{-2}$   which, as it is expected in crystals,
exhibits a pattern  of periodic order.
At low temperatures, the Lindemann's ratio $\gamma$ around the equilibrium
density tends 
to zero in classical solids while in   quantum crystals it is finite due to
the zero-point motion of particles, hence   this quantity is regarded
as a good quantum indicator. Furthermore, the Lindemann's ratio (or
equivalently, the mean squared displacement) is related to the Debye-Waller
factor $M_{Q}$~,  which describes the attenuation of the emergent radiation
in coherent scattering experiments according to the formula $ I\left(Q,T
\right) \propto e^{\left(-2M_{Q}\right)}$ (where  $I\left(Q,T\right)$ is
the intensity of the outgoing radiation scattered by the target and  $Q$ is
the modulus of the transfer wave vector). By means of a cumulant expansion,
the Debye-Waller factor can be expressed as  
\begin{equation}
\label{eq:debyexpans}
2M_{Q}=\langle u^{2}_{Q}\rangle Q^{2} - \frac{1}{12}\left(\langle u^{4}_{Q}\rangle 
-3\langle u^{2}_{Q}\rangle^{2}\right)
Q^{4}+O(Q^{6}) ~,
\end{equation}
where $\langle u^{2}_{Q}\rangle$ is the mean squared displacement along the
direction  $\widehat{{\bf Q}}$. It is easy to see that when the
distribution of particles around the equilibrium positions is
well-described by a Gaussian function, the quantity within parentheses in
the second right-term of Eq. (\ref{eq:debyexpans}), known as kurtosis $\zeta_{Q}$~,
vanishes. In such a case, the Debye-Waller factor reduces to the simple
formula $2M_{Q} = \langle u^{2}_{Q}\rangle Q^{2}$~. In
Table~III~, we report
the Lindemann's ratio, kurtosis and mean squared displacement of
two-dimensional p-${\rm H_{2}}$ at different densities. We have calculated
$\langle u^{2}_{Q}\rangle$ and $\zeta_{Q}$ along two orthogonal directions
and not found appreciable differences in the results. Moreover, the
kurtosis is null in all the studied cases. Consequently, we may conclude
that the distribution of hydrogen molecules around the equilibrium 
positions is isotropic and can be accurately reproduced by a Gaussian,
contrarily to what it is found to occur in
$^4$He.~\cite{draeger00} 
Regarding the value of $\gamma$, it can be said that solid H$_2$ is
\textit{less} quantum than $^4$He since $\gamma_{\rm H_{2}} \sim 0.18$ at 
$\rho_0$ whereas in solid helium $\gamma_{\rm He}\sim 0.24$ 
near melting.~\cite{cazorla08a}
Also it is worth noticing that the trend of $\gamma_{\rm H_{2}}$ is to increase
with decreasing density, therefore quantum exchange effects  in the crystal would 
become of greater relevance at small densities.

\section{Molecular ortho-deuterium}
\label{sec:d2}

The ground-state properties of o-${\rm D_{2}}$ (with total
angular momentum zero) have been also studied using the DMC method and the same  radial pair 
potential (Silvera-Goldman)  than in  p-${\rm H_{2}}$. 
The larger mass of D$_2$  makes one to expect that two-dimensional bulk ${\rm D_{2}}$
is solid at zero temperature, so in this case we have restricted our study
to the solid phase. 

\begin{figure}
\centerline{
        \includegraphics[width=0.8\linewidth]{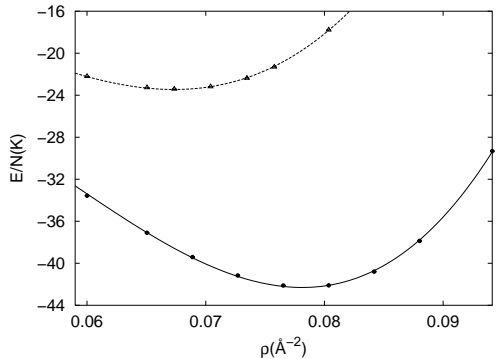}}%
        \caption{Ground-state energy per particle of 2D solid o-${\rm D_{2}}$ (solid line and filled circles).
and 2D solid p-${\rm H_{2}}$ (dotted line and empty triangles) which is shown for comparison. }
\label{fig:deuterium}
\end{figure}

In our simulations, the equilibrium
positions of the o-${\rm D_{2}}$ molecules are  arranged according to a
triangular lattice and the particles are assumed point-like. In this case,
we use the trial wave function  
\begin{equation}
\label{eq:deuterwf}
\psi_{NJ}'\left({\bf r}_{1},{\bf r}_{2},\ldots,{\bf r}_{N}\right)= \prod_{i<j}^{N}e^{-\frac{1}{2}\left(b/r_{ij}\right)^{c}}
\prod_{i}^{N}e^{-\frac{a}{2}\left(|{\bf r}_{i}-{\bf R}_{i}|\right)^{2}}~,
\end{equation}
which differs slightly from $\psi_{NJ}$ in Eq.~(\ref{eq:trialwvfuncsol})  
( now the pair correlation factors ${\rm f_{2}}$ depend on the extra
variational parameter $c$~). The variational parameters in Eq.
(\ref{eq:deuterwf}) have been optimized using VMC; the best values are 
$b = 3.32$ \AA~, $c = 7$, and $a =0.67$~\AA$^{-2}$~. 
 All the
DMC simulations have been performed in a rectangular plane containing 
$120$ particles and applying periodic boundary conditions.
 The target averaged population of walkers, $n_{w}$,
is $250$ and the time step, $\Delta\tau$,  $5 \cdot 10^{-4}$~K$^{-1}$~.
Finite size effects have been corrected with the same approach than used
for hydrogen (see Sec.~\ref{subsec:DMC}).

\begin{table}
\begin{center}
\label{tab:d2results}
\begin{tabular}{c c c c}
\hline
\hline
$ \rho$ (\AA$^{-2}$) & $E/N$ (K) & $V/N$ (K) & $T/N$ (K) \\
\hline
$   0.046 $ & $ -22.40 (2)$ & $ -29.84 (3) $ & $ 7.44 (3) $ \\
$   0.053 $ & $ -27.97 (1)$ & $ -37.70 (3) $ & $ 9.73 (3) $ \\
$   0.060 $ & $ -33.57 (1)$ & $ -46.38 (4) $ & $ 12.81 (4)$ \\
$   0.069 $ & $ -39.41 (1)$ & $ -57.52 (3) $ & $ 18.11 (3)$ \\
$   0.076 $ & $ -42.12 (1)$ & $ -66.46 (6) $ & $ 24.34 (6)$ \\
$   0.084 $ & $ -40.80 (1)$ & $ -72.59 (6) $ & $ 31.79 (6)$ \\
$   0.094 $ & $ -29.32 (2)$ & $ -73.18 (8) $ & $ 43.86 (8)$ \\
\hline
\hline
\end{tabular}
\end{center}
\caption{ Ground-state total and pure potential and kinetic energies per 
particle of 2D o-${\rm D_{2}}$ at several densities. }
\end{table}

In Fig.~\ref{fig:deuterium}~, we plot the total ground-state energy per
o-${\rm D_{2}}$ molecule as a function of the density; the solid line
represents the best fit to our data following the polynomial function expressed in 
Eq.~(\ref{eq:efit})~. The best value of the parameters are $B = 241(3)$~K, $C
= 324(10)$~K, $e_{0} = -42.305(5)$~K and $\rho_{0} =
0.0785(2)$~\AA$^{-2}$~,  which lead to a spinodal density $\rho_{S} =
0.0641(2)$~\AA$^{-2}$~. By comparing with p-${\rm H_{2}}$~, we show that
o-${\rm D_{2}}$ is denser at equilibrium and
appreciably more bounded (the total energy decreases substantially). The
 heavier mass of the o-${\rm D_{2}}$ molecules, makes the
solid to reduce its kinetic energy and mean squared displacement at any
density (see Tables~II~,~III~,~IV and V~), 
thus allowing the system to
increase its equilibrium density in order to take advantage of the
attractive interparticle interaction.

\begin{table}
\begin{center}
\label{tab:d2lidemann}
\begin{tabular}{c c c c}
\hline
\hline
$ \rho$(\AA$^{-2}$)   &  $\gamma_{\rm D_{2}} $  & $ \zeta_{(10)}$   &  $\zeta_{(01)}$    \\
\hline
$0.053$   &     $0.204(1)$  &  $0.00(1)$ &    $-0.06(2)$  \\
$0.060$   &     $0.187(1)$  &  $0.01(1)$ &    $0.00(1)$   \\
$0.069$   &     $0.160(1)$  &  $0.00(2)$ &    $-0.01(1)$  \\
$0.073$   &     $0.149(1)$  &  $-0.01(1)$&    $0.01(1)$   \\
$0.078$   &     $0.139(1)$  &  $0.00(1)$ &    $ 0.01(1)$  \\
$0.080$   &     $0.135(1)$  &  $-0.02(1)$&    $ 0.00(1)$  \\
$0.088$   &     $0.124(1)$  &  $0.00(1)$ &    $-0.03(1)$  \\
$0.094$   &     $0.117(1)$  &  $-0.02(1)$&    $ 0.02(1)$  \\
\hline
\hline
\end{tabular}
\caption{Lindemann's ratio $\gamma_{\rm D_{2}}$, and kurtosis $\zeta$
of two-dimensional o-${\rm D_{2}}$ at different densities near the equilibrium.}
\end{center}
\end{table}

Concerning the structural properties of 2D solid o-${\rm D_{2}}$~, we have
calculated the radial pair distribution function at the equilibrium density
(see Fig.~\ref{fig:deuteriumgr}~), and the Lindemann's ratio  and kurtosis
at different points (see Table~V~)~. As it is shown in
Fig.~\ref{fig:deuteriumgr}~, the peaks of the radial pair distribution
function of 2D o-${\rm D_{2}}$ are sharper and somewhat closer than in
molecular hydrogen at equilibrium since the density and degree of
localization of the particles are larger in the first case. This statement
is also corroborated by the results contained in Table~V~, where the
Lindemann's ratio is invariably some tenths smaller than in p-${\rm
H_{2}}$ at the same density. As a matter
of comparison, the Lindemann's ratio of o-${\rm D_{2}}$ at equilibrium is
about $1.3$ times smaller than that of p-${\rm H_{2}}$ and $1.7$ than in
two-dimensional $^4$He~. Furthermore, as it could be expected
from our previous study of 2D  p-${\rm H_{2}}$ (Table \ref{tab:lindemann}), 
the kurtosis in
two-dimensional o-${\rm D_{2}}$ is practically null in both the two
orthogonal directions for which it has been calculated.

\section{One-body density matrix and off-diagonal long range order}
\label{sec:possiblesuper}

A fundamental function in the study of quantum systems is the one-body
density matrix $\varrho_{1}({\bf r},{\bf r'})$, defined as
\begin{equation}
\varrho_{1}({\bf r},{\bf r'}) = \langle \Phi_{0} | 
\widehat{\psi}^{\dagger}({\bf r}) \widehat{\psi}({\bf r'}) | \Phi_{0} \rangle ~,
\label{eq:onebody}
\end{equation}
where $\widehat{\psi}({\bf r'})$ and $\widehat{\psi}^{\dagger}({\bf r})$
are, respectively, the field operators which destroy a particle from
position ${\bf r'}$ and create another at position ${\bf r}$, and $\Phi_{0}$
is the  ground-state wave function. In particular, a finite value for 
$\lim_{r \to \infty} \varrho_{1}(r)$  proves the existence of off-diagonal
long range order (ODLRO) in the system, the measure being the condensate
fraction $n_{0}$. In quantum Monte Carlo, the one-body density matrix
 can be estimated by averaging the coordinate operator
$ \psi_{T}({\bf r}_{1}+{\bf r},{\bf r}_{2},\ldots,{\bf r}_{N}) / 
\psi_{T}({\bf r}_{1},{\bf r}_{2},\ldots,{\bf r}_{N})$~. 
Here, we use extrapolated estimators for $\varrho_{1}$  
since the pure estimation relying on forward walking is only applicable to diagonal operators.
Moreover, in order to get consistent results we have required that the two extrapolated estimators
of the same accuracy, i.e.~, 
$\varrho_{1}(r)= 2\varrho_{1}^{mix}(r) - \varrho_{1}^{var}(r)$ and 
$\varrho_{1}(r)= \left(\varrho_{1}^{mix}(r)\right)^{2} /
\varrho_{1}^{var}(r)$~(where \emph{mix} means obtained with DMC and 
\emph{var} with VMC), coincide within the present statistical uncertainty.

\begin{figure}
\centering
       { \includegraphics[width=0.8\linewidth]{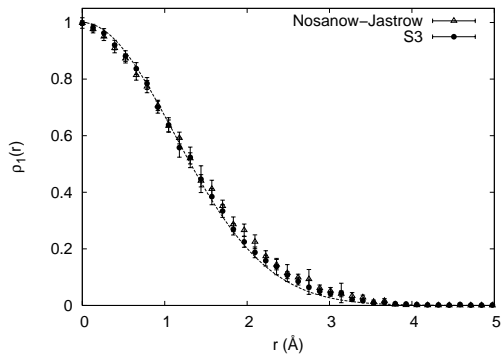} }%
\caption{One-body density matrix of p-${\rm H_{2}}$ $\varrho_{1} (r)$
obtained using as importance sampling $\psi_{NJ}$ and $\psi_{JG}^{S3}$
at the density $\rho = 0.060$~\AA$^{-2}$~. The solid line corresponds to
the Gaussian function which best fits to the Nosanow-Jastrow result.}
\label{fig:profile1}
\end{figure}

\begin{figure}
\centering
       { \includegraphics[width=0.75\linewidth]{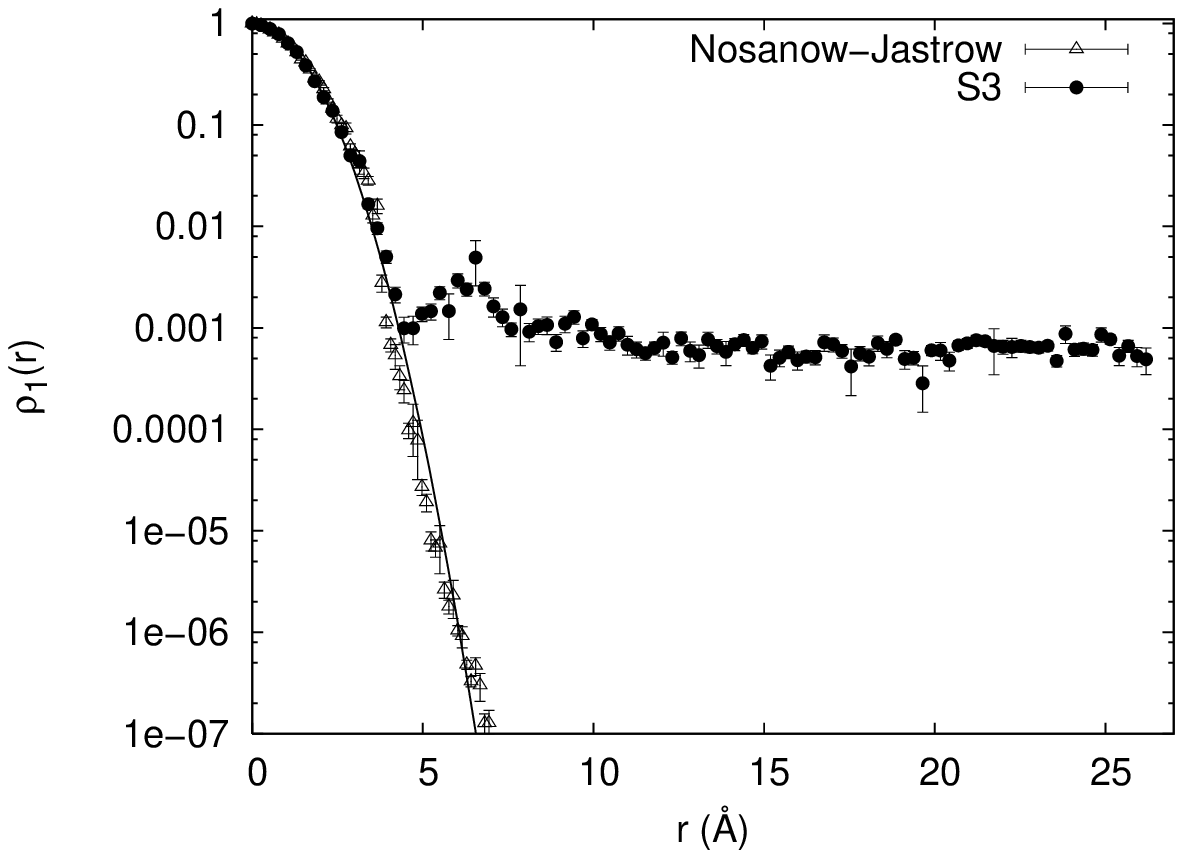} }%
       { \includegraphics[width=0.75\linewidth]{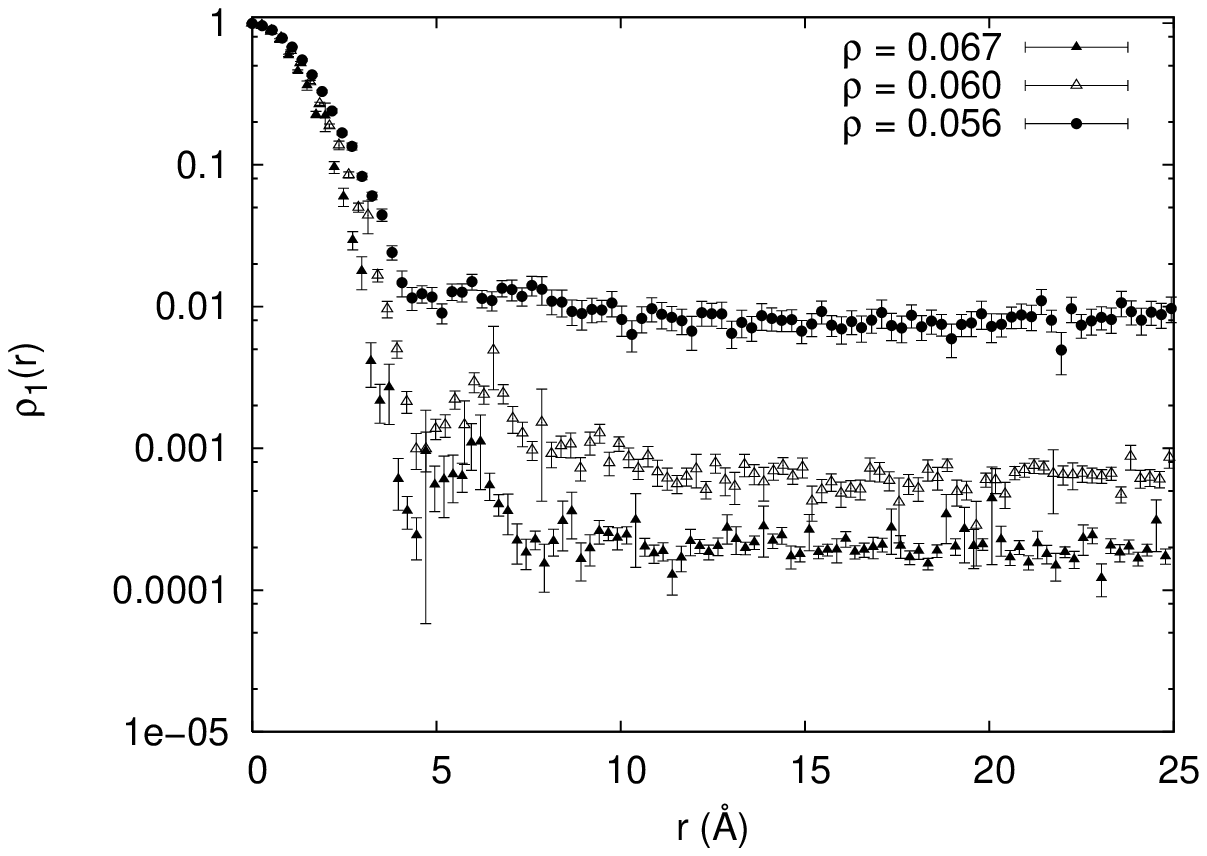} }%
\caption{\emph{Top}: One-body density matrix of p-${\rm H_{2}}$
at density $\rho = 0.060$~\AA$^{-2}$
obtained using importance sampling with the Nosanow-Jastrow and $\psi_{JG}^{S3}$ 
trial wave functions.
 \emph{Bottom}: One-body density matrix of p-${\rm H_{2}}$
obtained with the $\psi_{JG}^{S3}$ trial wave function at several densities. 
Densities are in units of \AA$^{-2}$~.}
\label{fig:condensate}
\end{figure}

In Fig.~\ref{fig:profile1}, we compare DMC results of the one-body density matrix
at the density $\rho = 0.060$~\AA$^{-2}$ and within the distance 
range $0 < r < 5.0$~\AA~, obtained with both $\psi_{NJ}$ (unsymmetrized) 
and $\psi_{JG}^{S3}$ (symmetrized) trial wave functions.
As one can observe therein, the series of points obtained with both twfs are compatible
in the full depicted range.
In the same graph, we also enclose the Gaussian curve $G(r) = e^{-br^{2}}$ which best fits to the
result obtained with $\psi_{NJ}$~;
we find in this case that the optimal value of the parameter $b$ is $0.400(6)$~\AA$^{-2}$~.
In order to test the quality of this fit (which in the reduced chi-squared test leads to the 
value $1.59$), we have calculated the atomic kinetic energy of two-dimensional p-${\rm H_{2}}$
through the formula
\begin{equation}
T/N = -\left[ \frac{\hbar^{2}}{2m_{\rm H_{2}}}\nabla^{2} \varrho_{1} (r)\right]_{r = 0} ~,
\label{eq:gaussianproof}
\end{equation}
but assuming $G (r)$ instead of $\varrho_{1} (r)$. 
In fact, it may be shown that Eq.~(\ref{eq:gaussianproof})
derives from the second moment of the momentum distribution $n(k)$,
\begin{equation}
T/N = \frac{\hbar^{2}}{2m_{\rm H_{2}}} \frac{1}{\left(2\pi\right)^{2}\rho}\int d{\bf k}~k^{2}~n(k)~.
\label{eq:sumrule}
\end{equation}
Proceeding in this way, we obtain $T/N = 19.40(30)$~K which does not agree satisfactorily with the corresponding
pure~(mixed) estimation $21.46(3)$~($20.75(2)$)~K~. 
Very interestingly, Withers and Glyde have recently
shown by means of simple models that anharmonic and/or particle-exchange effects in quantum solids may cause the 
momentum distribution $n ({\bf k})$, or equivalently $\varrho_{1} (r)$, to deviate significantly from a Gaussian 
function.~\cite{withers07} Therefore, on account of our variational results reported in Sec.~\ref{subsec:vmc}~ 
which show that molecule exchanges are likely to occur at very low rate, 
it may be suggested that two-dimensional hydrogen presents some degree of
anharmonicity.

In Fig.~\ref{fig:condensate}~(Top), we show DMC results similar to those enclosed in 
Fig.~\ref{fig:profile1} but for larger distances and expressed in logarithmic scale
in order to obtain the asymptote of $\varrho_1$~. 
As one observes in there, the value of $\lim_{r \to\infty} \varrho_{1}$ in the 
unsymmetrized case tends obviously to zero,
while for $\psi_{JG}^{S3}$ it amounts to a small but finite value
$n_0=6(1)\cdot 10^{-4}$~.  
In the same figure (Bottom), we compare $\varrho_{1}(r)$ at
several densities and only for the symmetric wave function;  we
obtain $n_0=2(1)\cdot 10^{-4}$ and $8(1)\cdot 10^{-3}$ at the density $0.067$
and $0.056$~\AA$^{-2}$~, respectively.   
Apart from the fact that we obtain a small but finite condensate fraction for solid
p-${\rm H_{2}}$ in all the studied cases, we note that the value of $n_{0}$
raises very abruptly  in moving from equilibrium to densities close to the
spinodal point (where $P < 0$). In a
very recent work, we have analyzed  the superfluid nature of solid $^{4}$He
at zero temperature by means of $\psi_{JG}^{S3}$~.~\cite{cazorla08c} In
that work, we showed that the superfluid fraction of bulk solid $^{4}$He lies
below  $1 \cdot 10^{-5}$, whereas a clear superfluid signal of
$\rho_{s}/\rho = 3.2(1) \cdot 10^{-3}$ appears in the  presence of 1~\% of
vacancies.  In the case of vacancies, we found that the condensate fraction
increased by roughly a  factor two with respect to that of the perfect
crystal configuration. According to this outcome, a significant increase of
$n_{0}$ in our simulations  might be identified to the appearance of
superfluidity in the system.

\begin{figure}
\centering
       { \includegraphics[width=0.75\linewidth]{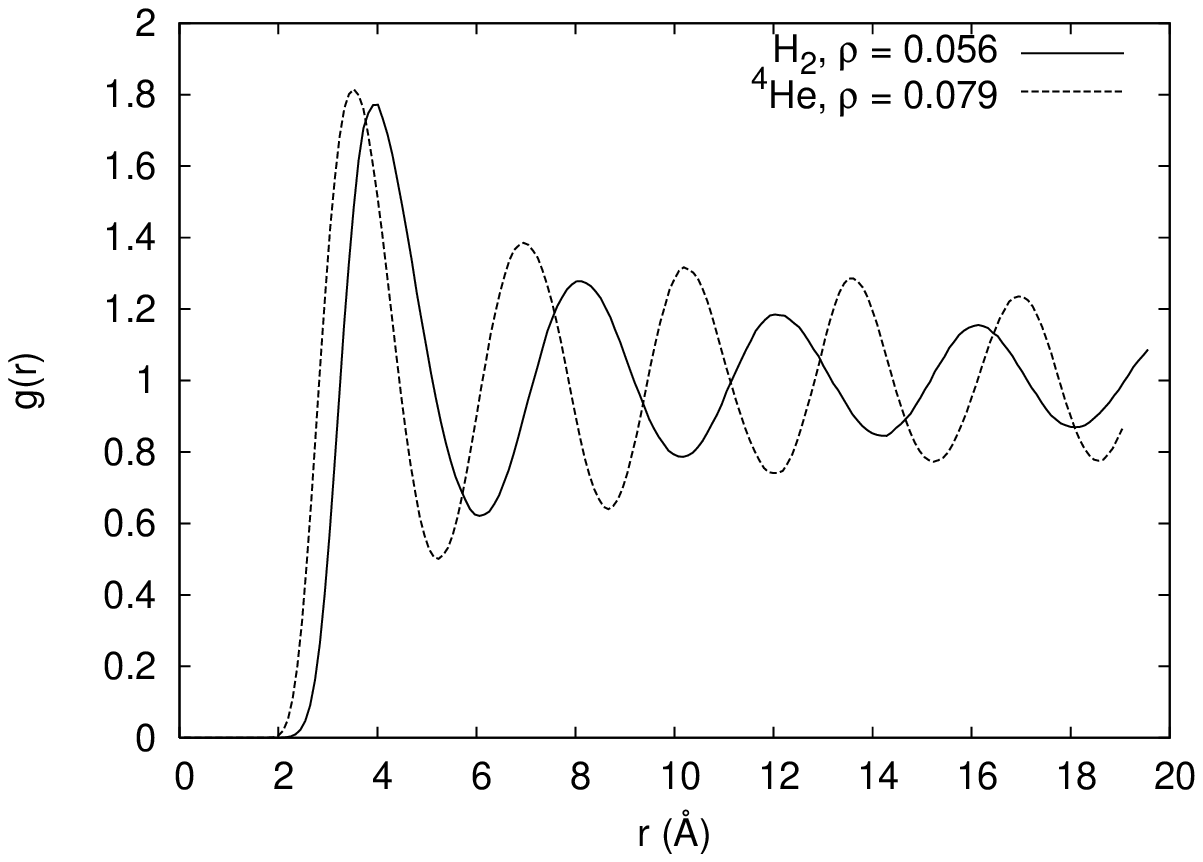} }%
       { \includegraphics[width=0.75\linewidth]{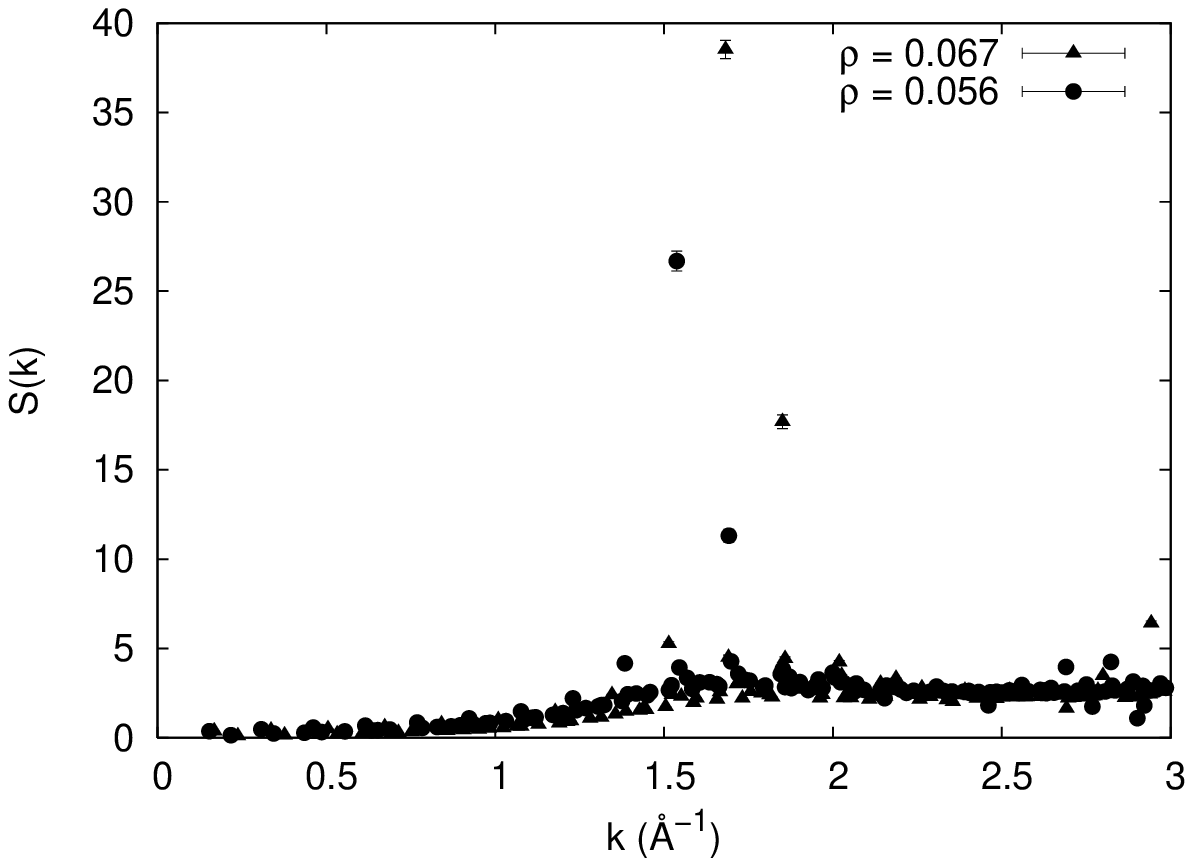} }%
\caption{
   \emph{Top}: Radial pair distribution function of two-dimensional ${\rm H_{2}}$ and $^{4}$He 
   near the  spinodal and freezing densities, respectively. \emph{Bottom}: Structure factor 
   $S(k)$ of 2D ${\rm H_{2}}$ at equilibrium density $\rho_0$ and 
   $\rho = 0.056$~\AA$^{-2}$~.}
\label{fig:compheh2}
\end{figure}

In oder to ensure that the significant raise of $n_{0}$ 
observed in our simulations is not due to partial melting of 
${\rm H_{2}}$, we have calculated the corresponding radial  pair
distribution function at the density $0.056$~\AA$^{-2}$~. In
Fig.~\ref{fig:compheh2}~(Top), we report $g(r)$ for hydrogen and
compare it with the one obtained for two-dimensional $^{4}$He above its
freezing point. Clearly, a typical solid pattern emerges for ${\rm H_{2}}$~.
Moreover, in the same figure (Bottom) we also plot the structure factor
$S(\rm k)$  for molecular hydrogen at
the equilibrium density $\rho_{0} = 0.067$~\AA$^{-2}$ and $\rho = 0.056$~\AA$^{-2}$~; 
in both cases marked peaks emerge at the reciprocal lattice vectors.
Therefore, the significant variation of $n_{0}$  that 
we observe when the  density decreases is not caused by  possible structural 
instabilities affecting the solid.

\section{Discussion and conclusions}
\label{sec:discussion}

To summarize, in this work we have studied two-dimensional p-${\rm H_{2}}$
and o-${\rm D_{2}}$ at zero-temperature and low pressures, with the
diffusion Monte Carlo method and the Silvera-Goldman semi-empirical pair
interaction. We have assessed several energetic and structural properties
of both systems, like the total and kinetic energy per particle, radial
pair distribution function, and  Lindemann's ratio and quoted so isotopic
quantum effects in hydrogen. Our results show that no stable liquid phase
exists and therefore reducing one dimension with respect to bulk it is not
enough to get the so-longly searched superfluid phase  of H$_2$.

Interestingly, Wiechert \emph{et al.} have reported very recently on an
experiment on molecular ortho-deuterium coadsorbed on graphite preplated by
a layer of Kr, up to temperatures of $\sim 1.5$~K.~\cite{wiechert04}  The
authors of this work claim evidence for the existence of a reentrant ${\rm
D_{2}}$ liquid at very low temperatures,  based on their heat capacity and
neutron diffraction measurements. The system explored by Wiechert \emph{et
al.} can be fairly modeled by a monolayer. However, on account of our
results for ${\rm H_{2}}$, the possibility of  pure two-dimensional liquid
deuterium at zero temperature  must be ruled out. 
 On the expectance of new and more explanatory
experiments, we may point that, assuming that thermal effects are
practically negligible, the role of the interactions between  the deuterium
molecules and the atoms of the substrate are the ones of relevance. 
Certainly, Turnbull and Boninsegni have already addressed recent work on
this direction  by means of the Path Integral Monte Carlo (PIMC) method and
simple interaction models.~\cite{boninsegni05,turnbull07} 
Further improvement on the modeling of coadsorbed systems, putting
especial emphasis on the description of the interactions and the effect of
corrugation with the substrate, may open new and challenging venues  for
the realization of superfluidity in p-${\rm H_{2}}$ and o-${\rm D_{2}}$
systems.~\cite{cazorla04}

At the variational level, we have analyzed
the quality of three different symmetrized trial wave functions based on
the  Nosanow-Jastrow model in describing 2D solid molecular hydrogen.
We have shown that the recently proposed  symmetrized wave function used
to describe the supersolid also characterizes hydrogen satisfactorily. By
using that wave function, we have studied the behavior of the one-body density
matrix of solid p-${\rm H_{2}}$ with density and predicted that the system
could become superfluid at very dilute densities (where $P < 0$). Further
work is being carried out to estimate the superfluid density in the
negative pressure region trying to confirm the signature observed in the
one-body density matrix.

\acknowledgments
We acknowledge financial support from DGI (Spain) Grant No. 
FIS2005-04181 and Generalitat de Catalunya Grant No. 2005GR-00779.

\end{document}